\documentclass[12pt,a4paper]{article} 
\usepackage{amsmath,amssymb,subcaption,pgfplots,float,graphicx}
\usepackage{epstopdf}
\usepackage{helvet}
\usepackage{authblk}
\usepackage[font=footnotesize,labelfont=bf]{caption}
\usepackage{setspace}
\usetikzlibrary{arrows}
\def\bm{\boldsymbol}
\usepackage{titling}
\pretitle{\begin{center}\large\bfseries}
\posttitle{\par\end{center}\vskip 0.5em}
\preauthor{\begin{center}\large\ttfamily}
\postauthor{\end{center}}
\predate{\par\normalsize\centering}
\postdate{\par\small\centering}

\title{Model-based image analysis of a tethered Brownian fibre for shear stress sensing}
\author{M.~T.~Gallagher$^{1,2,3,\dagger}$, C.~V.~Neal$^{1}$, K.~P.~Arkill$^{4,5}$, D.~J.~Smith$^{1,2,3}$}
\date{%
	\scriptsize{
	$^1$School of Mathematics, and $^2$Institute for Metabolism and Systems Research, University of Birmingham, Birmingham, UK. B15 2TT.\\
	$^3$Centre for Human Reproductive Science, Birmingham Women's and Children's NHS Foundation Trust, Birmingham, UK. B15 2TG.\\
	$^4$School of Medicine, University of Nottingham, Nottingham, UK. NG7 2UH.\\
	$^5$Biofisika Institute (CSIC UPV/EHU), and Research Centre for Experimental Marine Biology and Biotechnology, University of the Basque Country, Bilbao, Spain.\\
	$^\dagger$ Corresponding author.
	}}

\newcommand{\shr}{\dot{\gamma}}
\newcommand{\dir}{\bm{d}}
\newcommand{\dirdot}{\dot{\bm{d}}}
\newcommand{\ethe}{\hat{\bm{\theta}}}
\newcommand{\ephi}{\hat{\bm{\phi}}}
\newcommand{\gradD}{\bm{\nabla}_{\!\dir}\,}
\newcommand{\angvel}{\tilde{\bm{\omega}}}

\providecommand{\bra}[1]{\left(#1\right)}
\providecommand{\e}{\mathbf{e}}
\providecommand{\Pe}{\mathrm{Pe}}
\renewcommand{\epsilon}{\varepsilon}
\renewcommand{\leq}{\leqslant}

\begin{document}

	\maketitle
	
	\abstract{The measurement of shear stress acting on a biologically relevant surface is a challenging problem, particularly in the complex environment of, for example, the vasculature. While an experimental method for the direct detection of wall shear stress via the imaging of a synthetic biology nanorod has recently been developed, the data interpretation so far has been limited to phenomenological random walk modelling, small angle approximation, and image analysis techniques which do not take into account the production of an image from a 3D subject. In this report we develop a mathematical and statistical framework to estimate shear stress from rapid imaging sequences based firstly on stochastic modelling of the dynamics of a tethered Brownian fibre in shear flow, and secondly on novel model-based image analysis, which reconstructs phage positions by solving the inverse problem of image formation. This framework is tested on experimental data, providing the first mechanistically rational analysis of the novel assay. What follows further develops the established theory for an untethered particle in a semi-dilute suspension, which is of relevance to, for example, the study of Brownian nanowires without flow, and presents new ideas in the field of multidisciplinary image analysis.}

\section{Introduction}
\label{sec:intro}

The force per unit area exerted on a surface by a moving fluid, otherwise known as wall shear stress (WSS), plays an important role in many physical and biological systems. For example, the function and structure of endothelial cells \cite{reneman2006,fisher2001}, and the design of microfluidic systems \cite{nge2013,el2006}. While there exist several ways of measuring WSS directly \cite{lee2009,grosse2007,brucker2005} these methods are not suitable for measuring WSS in, for example, the vasculature, as they either require insertion of deformable micropillars ($ \sim 100~\mu\textrm{m} $ tall), or neglect to take into account biologically relevant aspects of the flow, for example the pulsatile nature of the flow in the vasculature which also contains fluid particulates and has complex geometries. There are also other biological factors limiting such flow methods; the viscosity of many fluids of interest is often not known, and can change with time, introducing additional error into calculations. We also know that cell surface macromolecules (for example the glycocalyx), can extend a distance $ > 0.5~\mu \mathrm{m} $ into the fluid meaning that surface effects become important and difficult to calculate. The current method for measuring WSS in the vasculature relies on measurement of the velocity gradient on the wall through bulk flow techniques such as micro-particle image velocimetry ($ \mu $PIV) \cite{poelma2008,sugii2005,smith2003}. However, due to the size of the particles needed to measure flow through blood vessels, Brownian effects become important which can introduce error in the measurement of velocities, and uncertainty in the location of the particles. In the present research we turn the Brownian motion of particles to our advantage; instead of needing to correct for such effects, the Brownian motion of a tethered rod is the measurement mechanism which underpins this work.

	To measure shear stress in the vasculature at the same place that an endothelial cell can detect requires a sensor that can respond to shear stress in the same location. We continue the development of sensor that can detect shear stress in microvessels as close as a few hundred nanometers from the cell membrane in real time in live animals. A biological microrod approximately $1~\mu$m in length, based on M13 bacteriophage (hereafter referred to as \textit{M13}), has recently been demonstrated to act as such a surface shear stress sensor \cite{lobo2015} through flow-induced changes to its tethered Brownian motion. The M13 is $ 7$~nm wide and $ \approx 900$~nm long, forming a semi-rigid `nanorod' which can be genetically engineered, or chemically modified to bind to fluorescent moieties, or antibodies. These monodisperse nano-particles have been used to produce several nanoscale devices including nanowires \cite{chen2013,ghosh2012}, and scaffolds for PCR \cite{carr2015}. 	Other methods using orientations of freely suspended nanorods have been employed by Kim et al \cite{kim2017}, where the real-time measurement of the collective orientation of nanorods has been used to measure local shear rate in microfluidic systems. The collective orientation of suspensions of nanorods have also been recently used to detect pathogenic bacteria through the shear alignment of virus particles and linear dichroism by Pacheco-G\'omez et al \cite{pacheco2011}. These characteristics have been used to generate an M13 construct that includes a collagen antibody covalently attached to one end, and decorated with more than $ 500 $ fluorophores along its length. This construct allows the M13 to bind at one end to a collagen coated slide, and be imaged using epi-fluorescent microscopy. It is this construct that we will focus on in this report.

	The framework for the modelling and measurement of WSS constructed in this report consists of two key steps: modelling the dynamics of a tethered Brownian fibre, and the extraction of experimental data through the use of \textit{model-based image analysis}. The modular nature of this framework will mean that it can be easily extended to investigate related problems in both micro-scale biology and areas where reliable, rational analysis of experimental image data is desired.
		
	Under no flow, the attached M13 oscillates randomly due to Brownian motion. As a flow is applied, however, the M13 movement is biased towards the direction of flow. It is this biasing behaviour which will allow us to calculate the wall shear stress due to an applied flow through the direct measurement of the M13 direction.	Data interpretation has so far been limited to phenomenological random walk modelling, and small-angle approximation to the resulting partial differential equations; however to apply the M13 quantitatively and to assess effects such as surface topography and variations in fibre length, it is valuable to model the underlying fluid dynamics of the tethered rod. We develop a mathematical framework for the rotational Brownian dynamics of a tethered M13, utilising rational mechanistic modelling to gain deep understanding about the behaviour of the M13 and its relationship to WSS. What follows is relevant to the established theory for an untethered particle in a semi-dilute suspension \cite{kim2017,strand1987}, and also to, for example, the recent study of Brownian nanowires without flow by Ota et al.\ \cite{ota2014}. 
		
	Due to the width of the M13 ($7$~nm) being much smaller than the wavelength of light used to excite the attached fluorophores ($561$~nm) the produced image is heavily diffracted and as such it requires work to calculate the exact location of the M13. Traditionally, deconvolution algorithms would be applied to such an image, either with a priori knowledge of how the light has been diffracted or without (blind deconvolution), several such schemes are available as packages in both ImageJ \cite{schindelin2015} and MATLAB \cite{matlab2017} as well as others. Current methods to do this often involve the use of `black box' processing algorithms. While these tools can be can be useful, and often provide good information, a lack of transparency can hinder interpretation, particularly in a context where statistical properties of the error are crucial, and as such can never give complete confidence in the results. Even when the details of such algorithms are known, they often rely on changing the image without any knowledge of what the image contains or how it was formed. To combat this we develop here the concept of \textit{model-based image analysis}. Using knowledge of the physics of image formation, including understanding of how optical effects such as diffraction of light occur, we construct a mathematical framework for the inverse problem of image formation; how, given an experimental image, we can calculate what originally formed the image by undoing the image formation process. As well as providing a rational framework for analysing images, model-based image analysis produces consistent results and can be applied to any experimental set up where the knowledge of the image formation is sufficiently well understood.
	
	In this report we combine work from the areas of synthetic biology and mathematical modelling, together with fluid dynamics and the concept of model-based image analysis to create a framework for the measurement of wall shear stress in biological systems. In the first part of this work we present the dynamics of a tethered Brownian fibre, and relate the angle distribution of the M13 in flow to the P\'eclet number, the ratio between Brownian and convective effects in the flow. We continue by introducing the concept of model-based image analysis and the inverse problem of image formation, and include algorithms for the automated processing of the experimental image data. The automated nature of the image processing, as well as allowing large amounts of data to be analysed, allows for the analysis of accuracy of the methods through large-scale simulations of data. Finally, we combine all these ideas to calculate the WSS for the flow. The principle will then be demonstrated on the experimental data of Lobo et al \cite{lobo2015}, providing the first mechanistically rational analysis of this novel assay. 

\begin{figure}[t]
	\centering
	\includegraphics[width=0.6\textwidth]{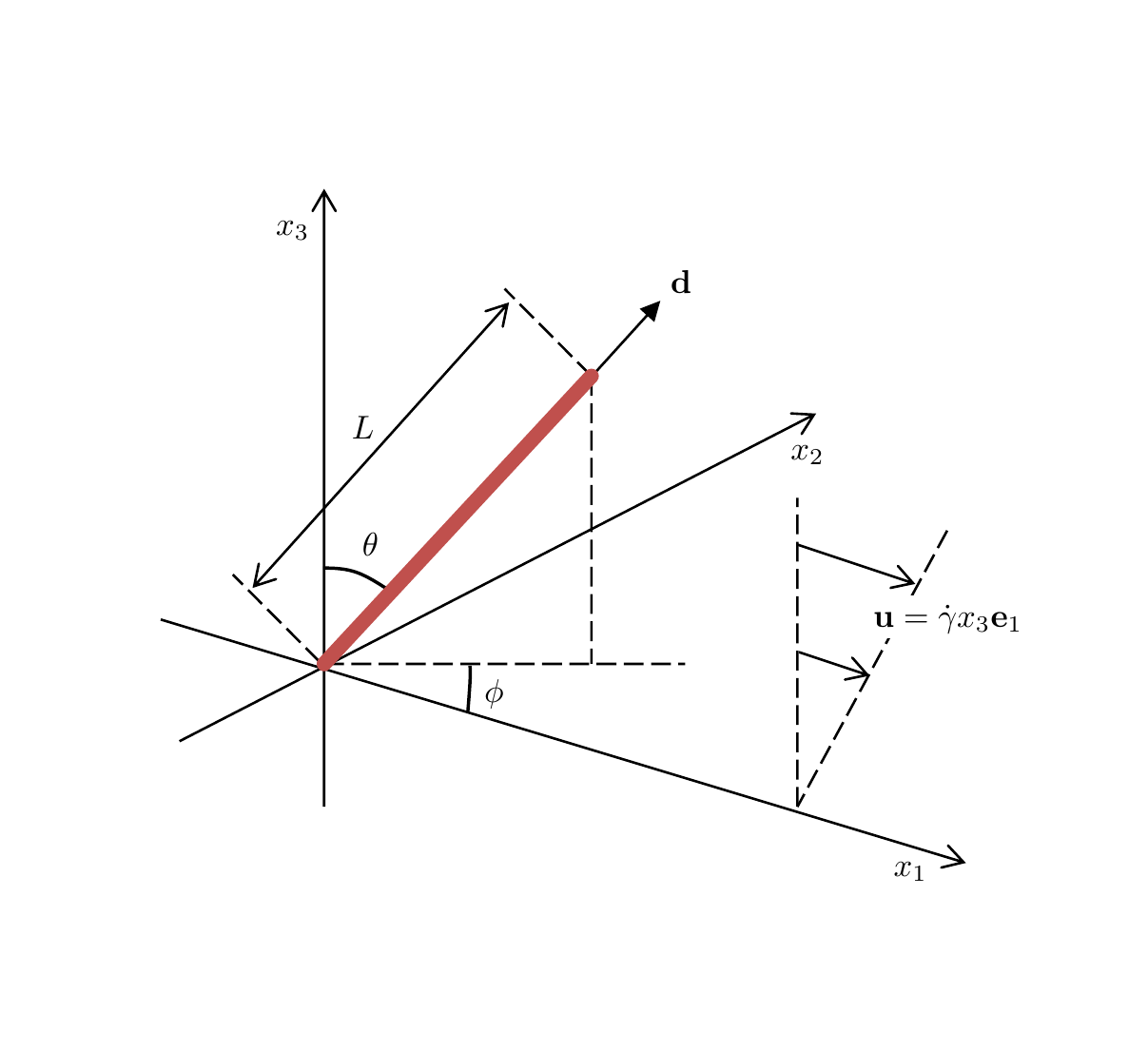}
	\caption{Definition sketch showing the location of the M13 (red), along with the direction of the applied shear flow. Here the M13 is tethered at $\bra{0,0,0}$ to the solid $ \bra{x_1,x_2} $-plane, with direction vector $ \mathbf{d} $.}
	\label{fig:def}
\end{figure}


\section{Dynamics of a tethered Brownian fibre}
\label{sec:pec}

We model the rotational Brownian dynamics of a rigid axisymmetric fibre of length $L$ projecting into the half-space $x_3>0$, attached at $(0,0,0)$ to the solid plane boundary $x_3=0$ under homogeneous unidirectional shear flow $\bm{u}=\shr x_3 \bm{e}_1$. A definition sketch is included in figure \ref{fig:def}. This choice of flow and geometry will provide a strong basis upon which these methods can be extended to reflect other interesting biological problems. Working in spherical polar coordinates $(r,\theta,\phi)$ and following Kim \& Karrila \cite{kim1991} we define $\dir(\theta,\phi)$ to be the direction vector of the M13, with the triple $\left[\dir,\ethe,\ephi\right]$ being the basis vectors. We denote by $\gradD$ and $\gradD \,\cdot$ the angular parts of the spherical polar gradient and divergence operators,
	\begin{align}
		\gradD f &= \partial_{\theta} f \ethe + \frac{1}{\sin\theta}\partial_{\phi} f \ephi \text{,}\\
		\gradD \cdot \bm{F} & = \frac{1}{\sin\theta}\partial_{\theta} (\sin\theta F_\theta) + \frac{1}{\sin\theta}\partial_\phi F_\phi.
	\end{align}
The problem will be to determine the steady state of the probability density function $\psi(\theta,\phi,t)$ for the fibre orientation, on the unit hemispherical domain $0^{\circ}\leq \theta \leq 90^{\circ}$ and $0^{\circ}\leq \phi < 360^{\circ}$\footnote{Here we have shown all angles in degrees for consistency with the results presented in Lobo et al \cite{lobo2015}. However all calculations have been performed in radians.}. The probability density will satisfy the normalisation condition,
\begin{equation}
	\int_{0^\circ}^{360^{\circ}} \int_{0^\circ}^{90^{\circ}} \psi(\theta,\phi,t) \sin\theta \, d\theta \, d\phi = 1. \label{eqn:normalisation}
\end{equation}
Note the change relative to \cite{strand1987,kim1991} in the absence of the $4\pi$ factor in equation~\eqref{eqn:normalisation}, so that the unscaled $\psi$ is a probability density function (the factor of $4\pi$ is less appropriate when working on a hemispherical domain). Two-dimensional imaging will directly yield a projection onto the $(x_1,x_2)$-plane, so we will observe samples from the marginal density function,
\begin{equation}
\Phi(\phi,t):=\int_{0^\circ}^{90^{\circ}} \psi(\theta,\phi,t) \sin\theta \, d\theta.
	\label{eqn:mPDF}
\end{equation}

The flux vector of $\psi$ in $ \bra{\theta,\phi} $ space is given by $\bm{J}=\psi \dirdot$, where $\dirdot(\theta,\phi)$ is the rate of change of $\bm{d}$ due to the combination of hydrodynamic and Brownian rotations. After some work we obtain the advection-diffusion equation
\begin{equation}
	\partial_t \psi + \gradD\cdot(\bm{\alpha}\psi)=\gradD\cdot (\mathcal{D}\gradD \psi),
	\label{eqn:AdvDiff}
\end{equation}
where $\mathcal{D}\bra{\theta}$ is the rotational diffusion matrix and $\bm{\alpha}\bra{\theta,\phi}$ is the rotational advection vector. Details of the derivation of \eqref{eqn:AdvDiff} are given in appendix \ref{app:der}. Introducing  dimensionless variables $t'$, $\mathcal{D}'$, $\bm{\alpha}'$ we have
\begin{equation}
t=\tau t' \mbox{,} \quad \mathcal{D}=\tau^{-1} \mathcal{D}' \mbox{,} \quad
\bm{\alpha}=\shr\bm{\alpha}
\end{equation}
with characteristic timescale $\tau=kT/\mu L^3$. The dimensionless advection-diffusion equation is then,
\begin{equation}
\partial_{t'} \psi + \gradD\cdot(\mathrm{Pe}\, \bm{\alpha}' \psi) = \gradD\cdot (\mathcal{D}'\gradD~ \psi)\mbox{,} 
\end{equation}
where the rotational P\'eclet number $\mathrm{Pe}=\shr \tau$. In the current work we make the assumption that $ \psi $ is independent of time for a given flow (for a fixed P\'eclet number), which gives the steady state dimensionless advection-diffusion equation 
\begin{equation}
	\gradD\cdot(\mathrm{Pe}\, \bm{\alpha}' \psi) = \gradD\cdot (\mathcal{D}'\gradD~ \psi).
\label{eqn:dimAdvDiff}
\end{equation}
The coefficients $\mathcal{D}'$ and $\bm{\alpha}'$ will be calculated by solving the dimensionless rotational resistance and mobility Stokes flow problems respectively, after which the probability density function $ \psi $ can be calculated by solving \eqref{eqn:dimAdvDiff} subject to the normalisation condition \eqref{eqn:normalisation}. We solve \eqref{eqn:dimAdvDiff} directly using a centred finite difference scheme in MATLAB \cite{matlab2017}. The full expression for \eqref{eqn:dimAdvDiff} is given in appendix \ref{sec:appAdvDiff}.
 
\subsection{Solution of the rotational resistance and mobility Stokes flow problems}

There exist several approaches to solving the resistance and mobility Stokes flow problems, including finite element, boundary integral and regularised stokeslet methods, in addition to approximations based on slender body theory. In this paper we apply a novel variation on the method of regularised stokeslets, namely the nearest-neighbour discretisation of Smith \cite{smith2017}. This method retains the `meshlessness' of the original formulation, with the added benefit of having a major reduction in computational cost.

The small Reynolds number associated with microscale flow justifies the use of the (dimensionless) Stokes flow equations,
\begin{equation}
-\bm{\nabla}p + \bm{\nabla}^2 \bm{u}=0 \mbox{,} \quad \bm{\nabla}\cdot \bm{u}=0 \mbox{,} \label{eq:stokes}
\end{equation}
where $p$ is pressure and $\bm{u}$ is velocity. The relevant boundary conditions are no-slip/no-penetration on the plane $\bm{u}(x_1,x_2,0,t)=0$, no-slip/no-penetration on the rigid body $\bm{u}(\bm{X},t)=\dot{\bm{X}}$, and convergence to a prescribed steady far-field flow $\bm{u}(\bm{x},t)\rightarrow \bm{u}^\infty(\bm{x})$ as $|\bm{x}|\rightarrow\infty$.

A solution to equation~\eqref{eq:stokes} with the given boundary conditions may be expressed as a regularised stokeslet boundary integral,
\begin{equation}
\bm{u}(\bm{x},t)=\iint_{S(t)} \bm{B}^\epsilon(\bm{x},\bm{X})\cdot\bm{f}(\bm{X},t) \, dS_{\bm{X}}+\bm{u}^\infty(\bm{x}). \label{eq:reginteq}
\end{equation}
$S(t)$ denotes the body surface, $f_k$ the hydrodynamic force per unit area exerted by the body on the fluid, and $B_{jk}^\epsilon$ the regularised `blakelet' found by Ainley et al \cite{ainley2008},

\begin{figure}[t]
	\centering
	\includegraphics[width=\textwidth]{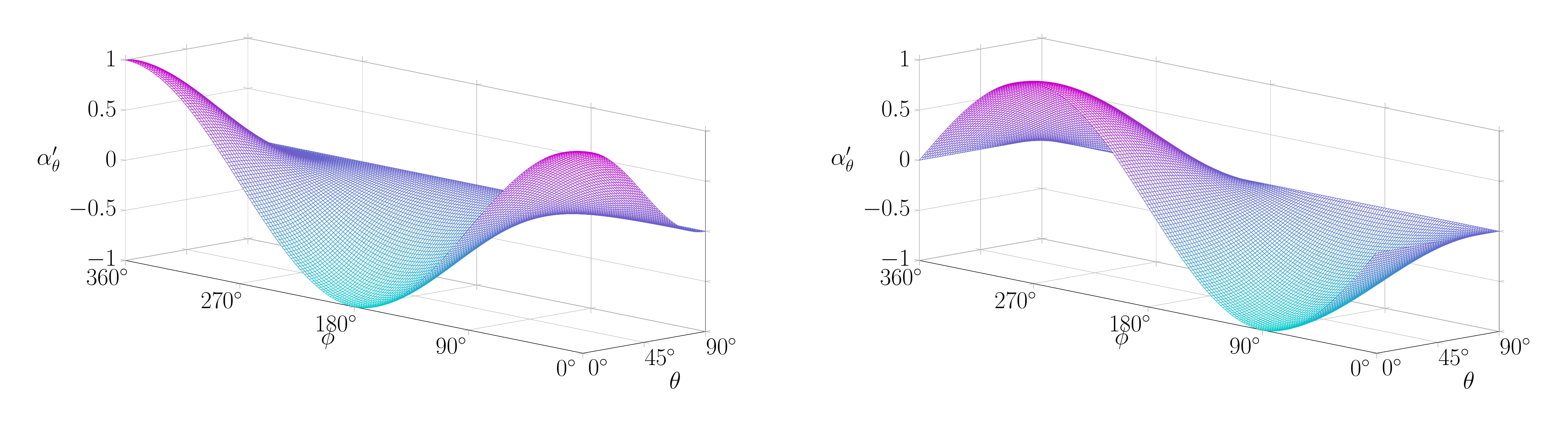}
	\caption{Components of the dimensionless rotational advection vector $ \bm{\alpha}^{\prime} $ plotted for $ 0^{\circ} \leq \phi < 360^{\circ} $, $ 0^{\circ} \leq \theta \leq 90^{\circ} $.}
	\label{fig:ppA}
\end{figure}

\begin{align}
B_{jk}^\epsilon(\mathbf{x},\boldsymbol{\xi}) =& \frac{1}{8\pi\mu}\bigg
    (\frac{\delta_{jk}(r^2 + 2\epsilon^2) + r_jr_k}{r_\epsilon^3} -
    \frac{\delta_{jk}(R^2 + 2\epsilon^2) + R_jR_k}{R_\epsilon^3}
    \nonumber \\
    +& 2h\Delta_{kl}\left[ \frac{\partial}{\partial R_l} \left(
    \frac{hR_j}{R_\epsilon^3} - \frac{\delta_{j3} (R^2 + 2\epsilon^2) + R_j
    R_3}{R_\epsilon^3} \right) - 4\pi h \delta_{jl} \phi_\epsilon(R)\right]
    \nonumber \\
    -& \frac{6h\epsilon^2}{R_\epsilon^5}(\delta_{j3}R_k -
    \delta_{jk}R_3)\bigg )\mbox{,}         \label{eq:regblakelet}
\end{align}
where $\epsilon$ is a small regularisation parameter, taken to be $ 1\% $ of the phage length.

Imposing the boundary conditions on the surface of the body, along with rigid body rotations about the origin, we have,
\begin{equation}
\bm{\omega}\times\bm{x}=\iint_{S(t)}\bm{B}^\epsilon(\bm{x},\bm{X})\cdot\bm{f}(\bm{X},t) \, dS_{\bm{X}}+\bm{u}^\infty(\bm{x})\mbox{,} \quad \mbox{for all} \quad \bm{x}\in S(t). \label{eq:angvelbie}
\end{equation}
In the inertialess regime, the system is closed by specifying the torque on the body due to hydrodynamic stress,
\begin{equation}
\bm{T}=\iint_{S(t)} \bm{X}\times\bm{f}(\bm{X},t)\,dS_{\bm{X}}. \label{eq:torquebal}
\end{equation}

The \emph{mobility problem} for this set-up then corresponds to the system of equations~\eqref{eq:angvelbie}--\eqref{eq:torquebal} with $\bm{T}$ and $\bm{u}^\infty$ prescribed and $\bm{\omega}$ unknown. The \emph{resistance problem} corresponds to the same system with $\bm{\omega}$ and $\bm{u}^\infty$ prescribed and $\bm{T}$ unknown.

The dimensionless rotational advection vector $\bm{\alpha}'$ is then given by solving the mobility problem for $\angvel$, prescribing $\bm{T}=0$ (corresponding to zero applied torque) and $\bm{u}^\infty=x_3\bm{e}_1$ (corresponding to unit shear flow). Then we have that $\bm{\alpha}'=\angvel\times \dir$. Recall that $\bm{\alpha}'=\bm{\alpha}'(\theta,\phi)$, therefore it is necessary to find an approximate solution over the domain $(\theta,\phi)\in [0^{\circ},90^{\circ}]\times [0^{\circ},360^{\circ})$.

The dimensionless diffusion coefficient $\mathcal{D}'$ is given by solving the resistance problems for $\bm{T}_\theta'$ and $\bm{T}_\phi'$, prescribing respectively $\bm{\omega}=\bm{e}_\theta$ and $\bm{\omega}=\bm{e}_\phi$ (corresponding to the two rotational modes), along with zero incident flow $\bm{u}^\infty=0$. Once these torques are found, the dimensionless resistance matrix in $(\theta,\phi)$ coordinates can be assembled as, \(\mathcal{R}'=\left(\bm{T}_\theta' | \bm{T}_\phi'\right)\); the dimensionless diffusion coefficient is then, $\mathcal{D}'=(\mathcal{R}')^{-1}$. Recall that $\mathcal{D}'=\mathcal{D}'(\theta)$; an approximate solution must therefore be found for all $\theta\in [0^{\circ},90^{\circ}]$, where, without loss of generality, we can set $\phi=0^{\circ}$.

\begin{figure}[t]
	\centering
	\includegraphics[width=\textwidth]{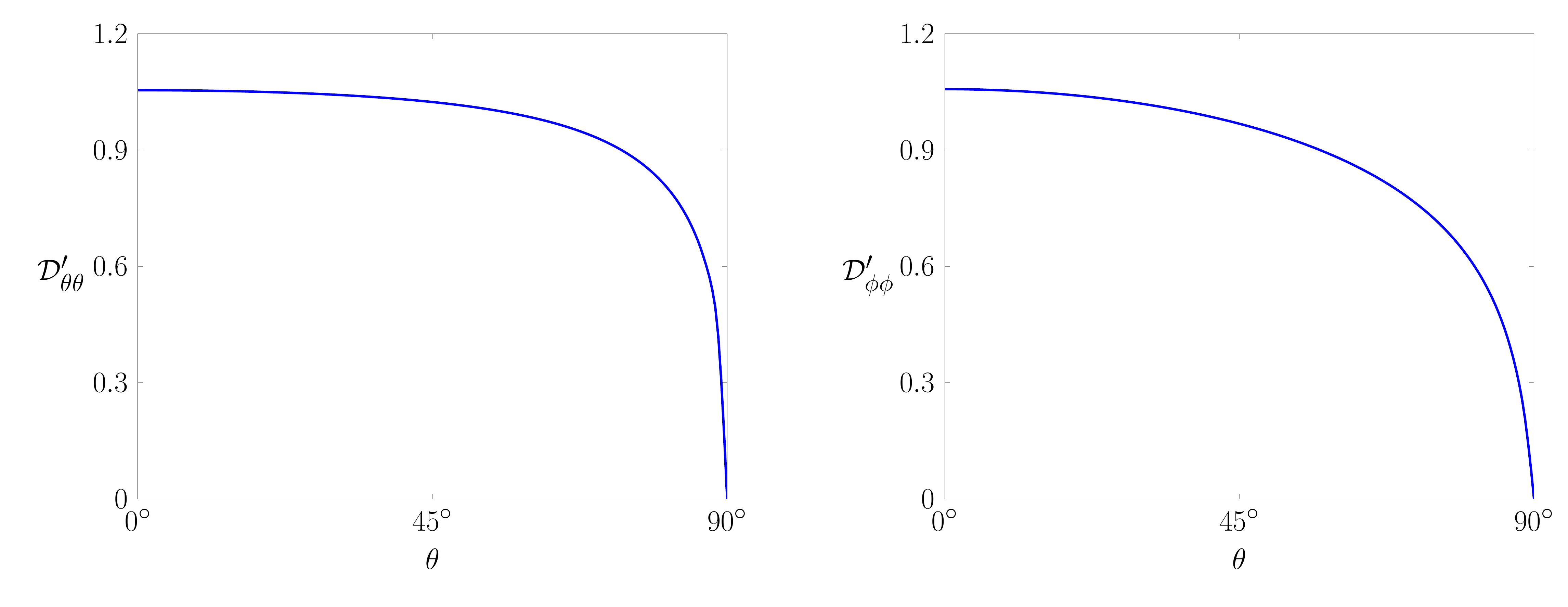}
	\caption{Non-zero components of the dimensionless diffusion matrix $ \mathcal{D}^{\prime} $ plotted against $ 0^{\circ} \leq \theta \leq 90^{\circ} $.}
	\label{fig:ppD}
\end{figure}

\subsection{Numerical Results}

The dimensionless rotational advection vector $ \bm{\alpha}^{\prime} $ is solved over a grid with $ 0^{\circ}\leq\theta\leq 90^{\circ} $ and $ 0^{\circ}\leq\phi\leq 360^{\circ} $, and is then interpolated using a cubic spline with periodic end conditions at the $ \phi $ limits. The resulting components $ \alpha_\theta $, and $ \alpha_\phi $ are shown in figure \ref{fig:ppA}. Similarly, the dimensionless rotational diffusion matrix $ \mathcal{D}^{\prime}\bra{\theta} $, is solved over $ {0^{\circ}\leq\theta\leq 90^{\circ}} $, is again interpolated using a cublic spline, and is shown in figure \ref{fig:ppD}. In solving for $ \mathcal{D}^{\prime} $ numerically we have introduced a small regularisation, at $ \theta = 90^{\circ} $, through enforcing $ \mathcal{D}^{\prime}\bra{90^{\circ}} = \delta $ (in our calculations we use $ \delta = 0.01 $). This ensures that the solutions for $ \mathcal{D}^{\prime} $ remain regular as $ \theta\rightarrow 90^{\circ} $. Finally, the advection-diffusion equation \eqref{eqn:dimAdvDiff} is solved for $ 1 \leq \Pe \leq 200 $. Here the bounds on $ \Pe $ have been chosen to include the experimentally relevant range for this project, but could be changed depending on the problem at hand. 

\begin{figure}[tp]
	\centering
	\includegraphics[width=0.45\textwidth]{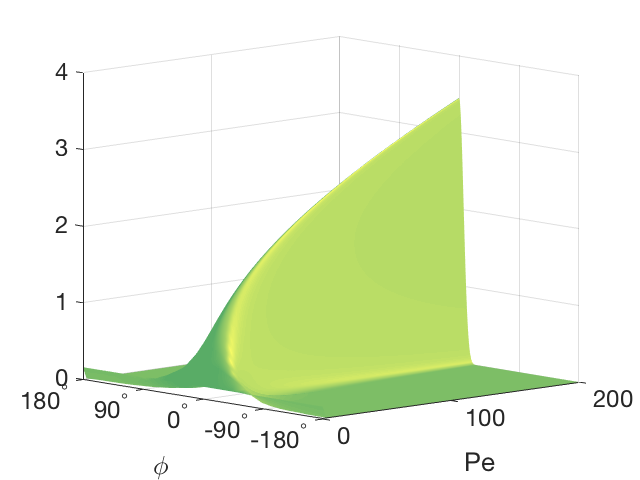}
	\includegraphics[width=0.45\textwidth]{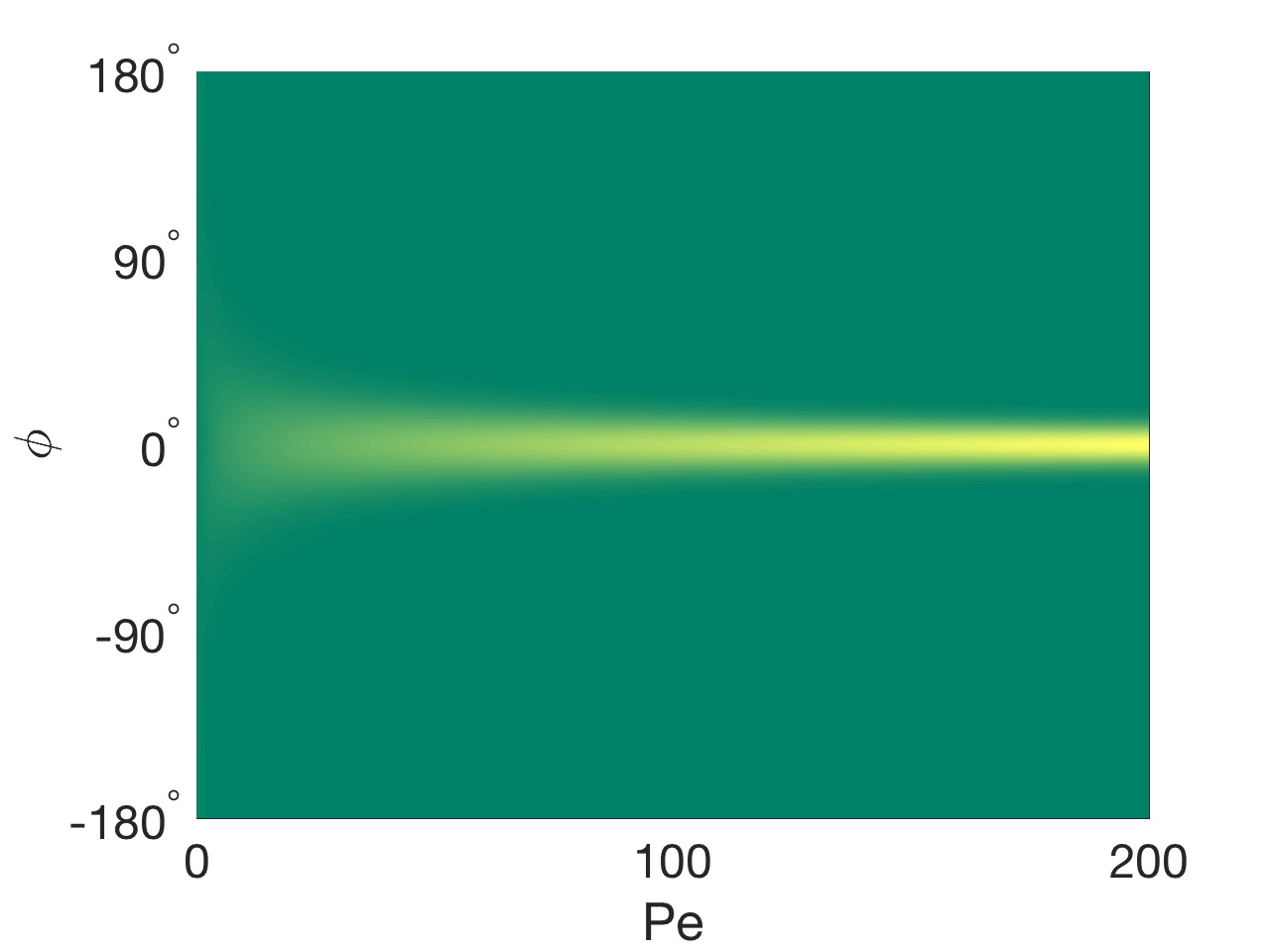}	
	\caption{Marginal probability density function $ \Phi $, the steady solution to the advection diffusion equation \eqref{eqn:mPDF}, plotted with ${ -90^{\circ} \leq\phi\leq 90^{\circ} }$, and $ 1 \leq \mathrm{Pe} \leq 200 $.}
	\label{fig:mpdf}	
\end{figure}

The marginal probability density function $ \Phi $ \eqref{eqn:mPDF} is obtained by integrating $ \psi $ over $ 0^{\circ}\leq\theta\leq 90^{\circ} $, the result is shown in figure \ref{fig:mpdf}. As expected, we see that the larger the P\'eclet number the more likely the M13 is to be aligned in the direction of the flow. Also as expected, when $ \Pe\rightarrow 0 $ we see the biasing effect decrease rapidly with the M13 approaching a uniform distribution. This behaviour is consistent with the physical interpretation of the P\'eclet number, with the case $ \Pe = 0 $ describing purely Brownian dynamics, with large P\'eclet numbers corresponding to shear dominated flows. Having calculated $ \Phi $ for a range of $ \Pe $ we should now able to estimate $ \Pe $ for a given set of angles $ \phi $. The methods by which we do this will be discussed in section \ref{sec:alg}.

In order to measure the WSS in a biological system given the theory presented above we require methods for the extraction of orientation data from experimental images. To this end we now turn our attention to developing the concepts of model-based image analysis.

\section{Detection of a tethered Brownian fibre}
\label{sec:meth}

	Having established a mathematical model for the dynamics of a tethered Brownian fibre, we now turn our attention to the application of the model to the experimental data of Lobo et al \cite{lobo2015}, with a view to calculating the P\'eclet number for an applied shear flow. The experimental procedure for obtaining images of the tethered M13 is contained within \cite{lobo2015} and as such not repeated here, except for noting that the experimental set up was that of a fluorescently labeled M13 tethered to a collagen coated slide which was then imaged with a $ 1.4 $NA oil objective with a spinning disk confocal microscope (Ultraview, PerkinElmer). In what follows we attack the problem through novel mathematical model based image analysis methods which, along with the theory presented in section \ref{sec:pec}, will provide a more rigorous and extensible basis for future work.

\subsection{The inverse problem of image formation}
\label{sec:inv}

We model the experimental M13 as a rigid, inextensible, axisymmetric rod of length $ L $ projecting into the half-space $ x_3 \geq 0 $. The M13 is tethered at the point $ \bra{x_0,y_0,z_0} $ of the Cartesian coordinate system $ \bra{x_1,x_2,x_3} $ to the solid plane boundary $ x_3 = 0 $, and is subjected to homogeneous unidirectional shear flow $ \mathbf{u} = \dot{\gamma} x_3 \e_1 $. The position of the M13 $ \bra{x_1,x_2,x_3} = \bra{x,y,z} $ is then given, in spherical polar coordinates, as
\begin{equation}
	x\bra{s} = x_0 + s\sin\theta\cos\phi,\quad y\bra{s} = y_0 + s\sin\theta\sin\phi,\quad z\bra{s} = z_0 + s\cos\theta,
	\label{eqn:M13}
\end{equation}
for given azimuthal and polar angles $ 0^{\circ}\leq \theta \leq 90^{\circ} $ and $0^{\circ}\leq\phi < 360^{\circ}$, with $ 0 \leq s \leq L $ being arclength along the M13. See figure \ref{fig:def} for a sketch of the setup noting that, in what follows, we now model the M13 as being tethered to some, as yet, unknown point $ \bra{x_0,y_0,z_0} $.

  Following Zhang et al \cite{zhang2007}, we model the optical diffraction of a light source located at the point $ \bra{X_0,Y_0,Z_0} $, diffusing over the focal plane $ \bra{X,Y,Z} $, by a Gaussian point spread function (PSF), namely
\begin{equation}
	P\bra{X,Y,Z} = I \exp\bra{-\frac{\bra{X-X_0}^2}{2\sigma_x^2} -\frac{\bra{Y - Y_0}^2}{2\sigma_x^2}-\frac{\bra{Z -Z_0}^2}{2\sigma_z^2}},
	\label{eqn:psf}
\end{equation}
where $ I $, $ \sigma_x $, and $ \sigma_z $ are parameters relating to the experimental setup. Note that we have assumed that the optical diffraction will be equal in both the $ \e_1 $ and $ \e_2 $ directions when imaged from above, resulting in a circular PSF for a given focal plane $ z = z_0 $. The resulting image, $ \mathbf{I} $, given by convolution of the point spread function (\ref{eqn:psf}) with the M13 location (\ref{eqn:M13}), in the focal plane $ \bra{x_1,x_2,0} $, is then
\begin{align}
	\mathbf{I}\bra{x_1,x_2} &= \mathbf{B}\bra{x_1,x_2}\notag\\
	&\quad + \int\limits_{0}^{L} I \exp\bra{-\frac{\bra{x_1-x\bra{s}}^2}{2\sigma_x^2} -\frac{\bra{x_2 - y\bra{s}}^2}{2\sigma_x^2}-\frac{z\bra{s}^2}{2\sigma_z^2}} \mathrm{d}s,
	\label{eqn:image}
\end{align}
where $ \mathbf{B} $ is some background image intensity, which may be constant or may vary with pixel location.

Given a set of experimental images, and a model for the forward problem of image formation (\ref{eqn:image}), it remains to solve the inverse problem of image formation; estimation of the position of the M13 given an experimental image. In order to ensure good fit between the experimental and simulated images, we choose the intensity parameter $ I $ to be
\begin{equation}
	I = \max\bra{\frac{E_i - B_i}{\int\limits_{0}^{L}\exp\bra{-\frac{\bra{x_{1i}-x\bra{s}}^2}{2\sigma_x^2} -\frac{\bra{x_{2i} - y\bra{s}}^2}{2\sigma_x^2}-\frac{z\bra{s}^2}{2\sigma_z^2}} \mathrm{d}s}},
\end{equation}
over all $ i $ pixels in the image. We define the M13 location to be the set of spatial parameters $ \bra{x_0,y_0,\phi,\theta} $, and optical parameters $ \left(I,\sigma_x,\sigma_z,\right.$ $\left.\text{and}\ \mathbf{B}\right) $ which minimise the sum-squared error between the experimental and simulated images, namely
\begin{equation}
	S = \sum\limits_{i}\bra{E_i - I_i}^2,
\end{equation}
where $ E_i $ and $ I_i $ are the $ i^{\text{th}} $ pixels in the experimental and simulated images respectively. The minimisation is performed globally using the multilevel coordinate search algorithm, routine \textit{e05jb}, from the NAG Toolbox for MATLAB \cite{nag}, with a set of bounds on each of the parameters. Due to the complexity of the problem, and the lack of detailed information regarding the optical parameter $ \sigma_z $, in what follows we model each image as though it contains a M13 of variable projected length $ L $, inclined at an angle $ \theta = 90^{\circ} $ to the vertical. Here we constrain the M13 parameters through requiring $ \bra{x_0,y_0} $ to lie within the image, $ 0 < L < 2~\mu m $, and $ -90^{\circ} \leq \phi \leq 90^{\circ} $. We then require that the optical parameters have the following constraints: $ 0 \leq \left\lvert \mathbf{B} \right\rvert $, and $ \overline{\sigma} / 10 \leq \sigma_x \leq 10~\overline{\sigma} $, where $ \overline{\sigma} $ is given by following Zhang et al \cite{zhang2007}.

\subsection{Fitting procedure}
\label{sec:alg}

	In refining our fitting algorithms we found that a small amount of preprocessing of the experimental images led to a significant increase in the accuracy of the fits. The preprocessing step involves applying a $ 5\times5 $-pixel median filter \cite{arce2005} to the experimental image, followed by subtracting the median image intensity from all pixels in the image, and finally setting the values of all pixels with negative intensity to zero. The effect of this preprocessing step is analysed in section \ref{sec:sens}. We then perform a multi-stage fit in order to find the M13 and optical parameters which can best replicate the given experimental image as follows:
	
	\begin{itemize}
		\item We first fit the spatial parameters for initial optical parameters $ \sigma_{x} $ and $ \mathbf{B} $. Due to the preprocessing of the experimental images we choose $ \mathbf{B} = 0$. The PSF spread $ \sigma_x $ is approximated by following Zhang et al \cite{zhang2007} for the experimental setup.
		\item Having calculated a first guess for the spatial parameters, the value for $ \sigma_x $ is then fit, keeping all other parameters fixed. While, theoretically, the value of $ \sigma_x $ should be constant for all images from a given experiment, due to the preprocessing step, we allow some variation in $ \sigma_x $ to take place.
		\item The spatial parameters are now refit using the updated value for $ \sigma_x $.
		\item We then fit the image background $ \mathbf{B} $, while allowing a small change in $ \sigma_x $ if necessary.
		\item Finally, the new values of $ \sigma_x $ and $ \mathbf{B} $ are used to fit the spatial parameters $ \bra{x_0,y_0,L,\phi} $.
	\end{itemize} 
	
	Once the M13 and optical parameters have been obtained for all the experimental images we can use the theory discussed in section \ref{sec:pec} to estimate the P\'eclet number for a particular flow. Using the marginal probability density function for the flow $ \Phi $ (shown in figure \ref{fig:mpdf}), we can integrate to find the related cumulative density function (CDF) $ F_1 $, which can then be compared to the sample CDF $ F_2 $ through calculation of the  Kolmogorov-Smirnov statistic $ D $ \cite{daniel1990},
	\begin{equation}
		D = \sup_{-90^{\circ}~\leq~ \phi~\leq~ 90^{\circ}} \left\lvert F_1\bra{\phi} - F_2\bra{\phi}\right\rvert.
	\end{equation}
	The P\'eclet number $ \overline{\Pe} $ which minimises $ D $ is then chosen as the fit. This optimisation procedure is again done with the multilevel coordinate search algorithm ($e05jb$) from the NAG Toolbox for MATLAB \cite{nag}. The accuracy of the fitting procedures is now investigated.

\section{Accuracy of fluid dynamics modelling with model-based image analysis }
\label{sec:sens}

	In order for this model-based image analysis framework to be useful, it must be able to accurately fit the location of a series of M13, and the P\'eclet number corresponding to the flow over such M13. We investigate the accuracy of the fit by dividing the problem into two areas where error can be introduced, namely the image processing stage, and the calculation of WSS from a sample of orientation data. For each of these steps we will generate $ 180 $ sample images for a spread of P\'eclet numbers $ 1 \leq \Pe\leq 200 $, which is comparable to both the number of images and the flow rates of the associated experiments.
	
\subsection{Step 1: Error associated with image processing}
\label{sec:case1}

\begin{figure}[tp]
	\centering
	\includegraphics[width=\textwidth]{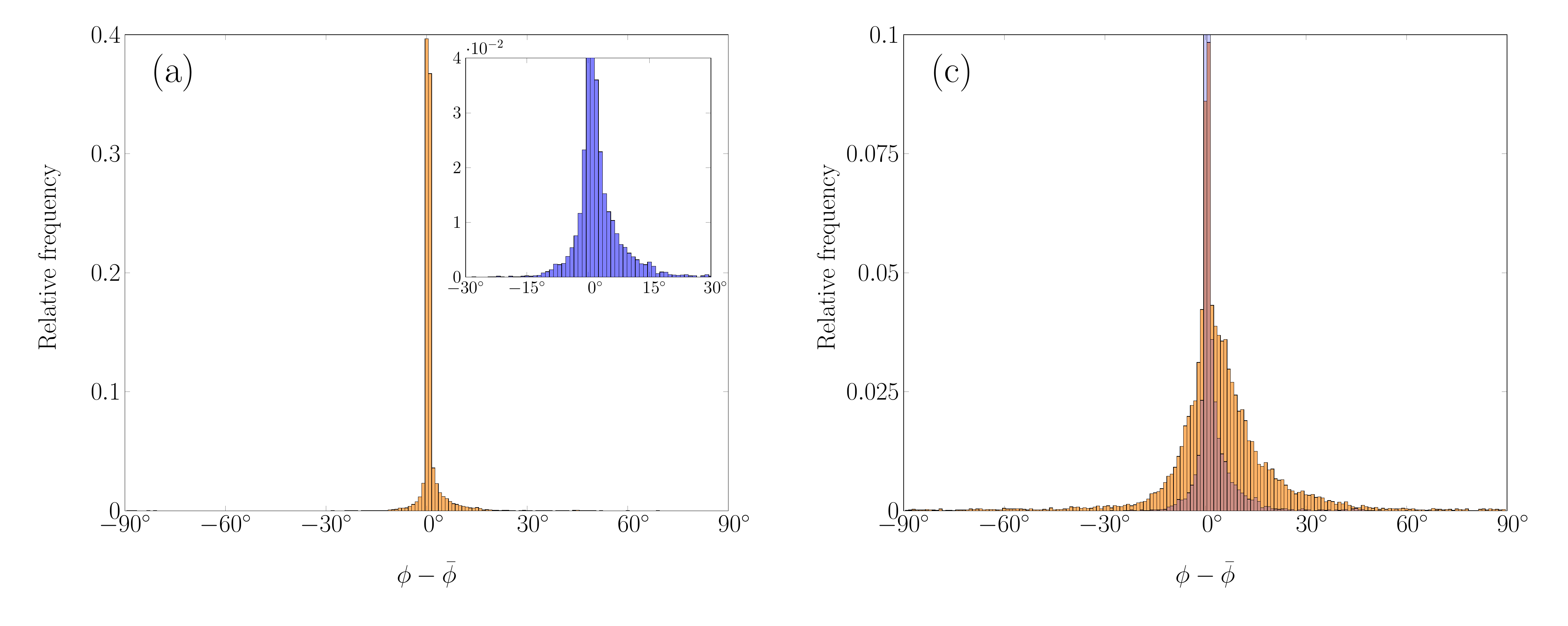}
	\caption{Relative frequency histograms of the error between the simulated M13 angles $ \phi $ and fit angles $ \bar{\phi} $, with $1^{\circ}$ bin widths, for analysis of the image fitting procedure. Figure (a) shows the error distribution with the preprocessing step included, with the insert (b) being the same figure zoomed in for clarity. Figure (b) shows the error distribution without the preprocessing step in orange, with the error with preprocessing  overlaid in blue.}
	\label{fig:ipHist}	
\end{figure}

\begin{figure}[t]
	\centering
	\includegraphics[width=\textwidth]{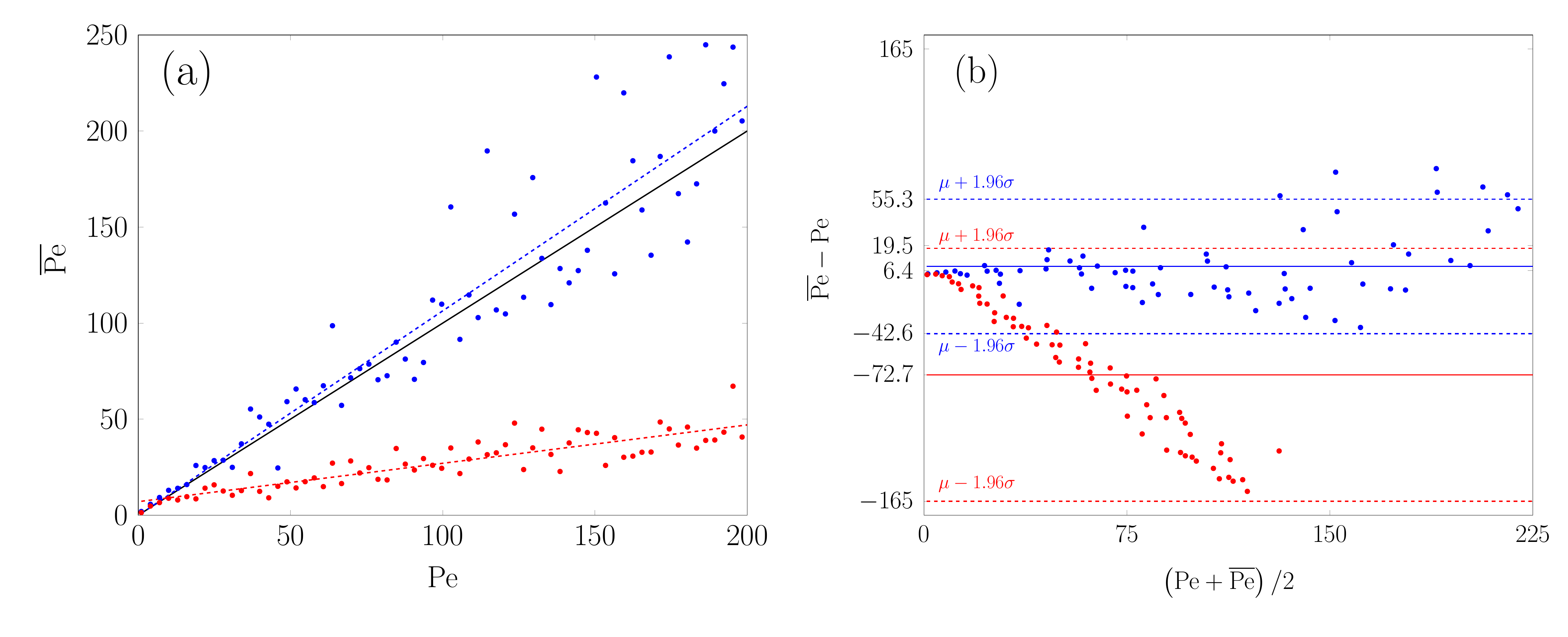}
	\caption{(a) plots the fit P\'eclet number $ \overline{\Pe} $ against simulated P\'eclet number $ \Pe $, for analysis of the image fitting procedure. The blue dots show the fit with the preprocessing step, while the red dots show the fit without. The black line shows what would be perfect correspondence between $ \Pe $ and $ \overline{\Pe} $, with the dotted lines being the line of best fit to the data. (b) shows the Bland-Altman plot testing the fit data $ \overline{\mathrm{Pe}} $ against $ \Pe $ with preprocessing (blue) and without (red). The solid lines show the mean of the difference between $ \overline{\Pe} $ and $ \Pe $ for each case, with the dotted lines being the $ 95\% $ confidence interval for the difference.}
	\label{fig:ipPe}	
\end{figure}

	In investigating the error associated in the image processing step, both with and without preprocessing, we would like to have a set of sample orientation data which, when fit, return the P\'eclet number corresponding to the distribution they were sampled from. To ensure this we use rejection sampling from the marginal PDF $ \Phi $ at a selection of linearly spaced P\'eclet numbers $1 \leq \Pe \leq 200$, stopping when we have a set of angles $ \phi $ which, when fit, give a P\'eclet number $ \overline{\Pe} $ such that $ \left\lvert\overline{\Pe} - \Pe\right\rvert < 0.5$. For each of these sets of angles a M13 is then simulated with a given length $ L $, and is placed at a point $ \bra{x_0,y_0} $, randomly chosen with $ -0.5$~$\mu$m $\leq x_0,y_0 \leq 0.5$~$\mu$m. An image of the M13 is then generated via \eqref{eqn:image}, with $ \sigma_x $ given by following \cite{zhang2007}. The intensity parameter $ I $ is chosen so that the image has a maximum intensity of $255$, which corresponds to the maximum value a 8-bit unsigned integer can take, and hence the maximum intensity in the experimental images. The additive noise $ \mathbf{B} $ in \eqref{eqn:image} is simulated by sampling from a normal distribution with a mean of $ 76.5 $ $ \bra{30\%\ \text{of}\ I} $, and a standard deviation of $ 5 $. These images are then put through both the image and P\'eclet fitting procedures, after which we are able to compare both the fitted angles $ \bar{\phi} $ and fit P\'eclet numbers $ \overline{\Pe} $. In order to evaluate the effectiveness of the preprocessing step, we analyse the same set of images twice, with and without the preprocessing step, and compare the results.

	The number of images successfully analysed and the number of fit orientation angles $ \bar{\phi} $ within $ 1^{\circ} $ and $5^{\circ}$ of simulated angles $ \phi $ is shown in table \ref{tab:ipErr}, with the corresponding relative frequency histograms of the error between the simulated angles $ \phi $ and fit angles $ \bar{\phi} $ are shown in figure \ref{fig:ipHist}. It is clear looking at this data that the inclusion of the preprocessing step improves the accuracy of the fit significantly. The P\'eclet numbers $ \overline{\Pe} $ obtained through analysis of the fit orientation angles $ \bar{\phi} $ is then shown in figure \ref{fig:ipPe}. We see here that not including the preprocessing step results in a significant under estimation of the P\'eclet number for the flow, while the inclusion of the preprocessing step leads to results which accurately represent the simulated flows. We see from the least squares line of best fit that the image processing method provides good results, with a small increase in error for stronger flows (higher P\'eclet number). This is in agreement with the Bland-Altman plot, figure \ref{fig:ipPe}b. Here we see a mean difference between $ \Pe $ and $ \overline{\Pe} $ of $ 6.4 $ with the preprocessing step, and $ 72.7 $ without. Similarly the standard deviation for the difference is $ 25 $ with preprocessing, compared to $ 47 $ without.

	\begin{table}
		\centering
		\caption{Table showing the number and percentage of successfully fit frames through the model-based image analysis procedure, and the number and percentage of fit angles $ \bar{\phi} $ within $1$ and $5$ degrees of the simulated angles $ \phi $. Here Step 1 and Step 2 correspond to the analyses in sections \ref{sec:case1} and \ref{sec:case2}. A total of $ 12060 $ images were analysed.}
		\label{tab:ipErr}
		\scalebox{0.8}{
		\renewcommand{\arraystretch}{1.2}{
		\begin{tabular}{| c | c || c | c | c |}
		\hline
			Data set & Method & Successfully fit & $\left\lvert\phi - \bar{\phi}\right\rvert < 1^{\circ}$ & $\left\lvert\phi - \bar{\phi}\right\rvert < 5^{\circ}$\\
			\hline\hline
			Step 1 & No preprocessing & $11674$ $\bra{96.80\%}$  & $2337$ $\bra{20.02\%}$  & $5684$ $\bra{48.69\%}$ \\
			\hline
			Step 1 & Preprocessing & $11766$ $\bra{97.56\%}$ & $9260$ $\bra{78.70\%}$ & $10840$ $\bra{92.13\%}$ \\
			\hline
			Step 2 & Preprocessing & $11736$ $\bra{97.31\%}$ & $9227$ $\bra{78.62\%}$ & $10779$ $\bra{91.85\%}$ \\
			\hline
		\end{tabular}}}
	\end{table}
 
\subsection{Step 2: Error associated with the full analysis of WSS from a sample of orientation data}
\label{sec:case2}
	
	\begin{figure}[t]
	\centering
	\includegraphics[width=\textwidth]{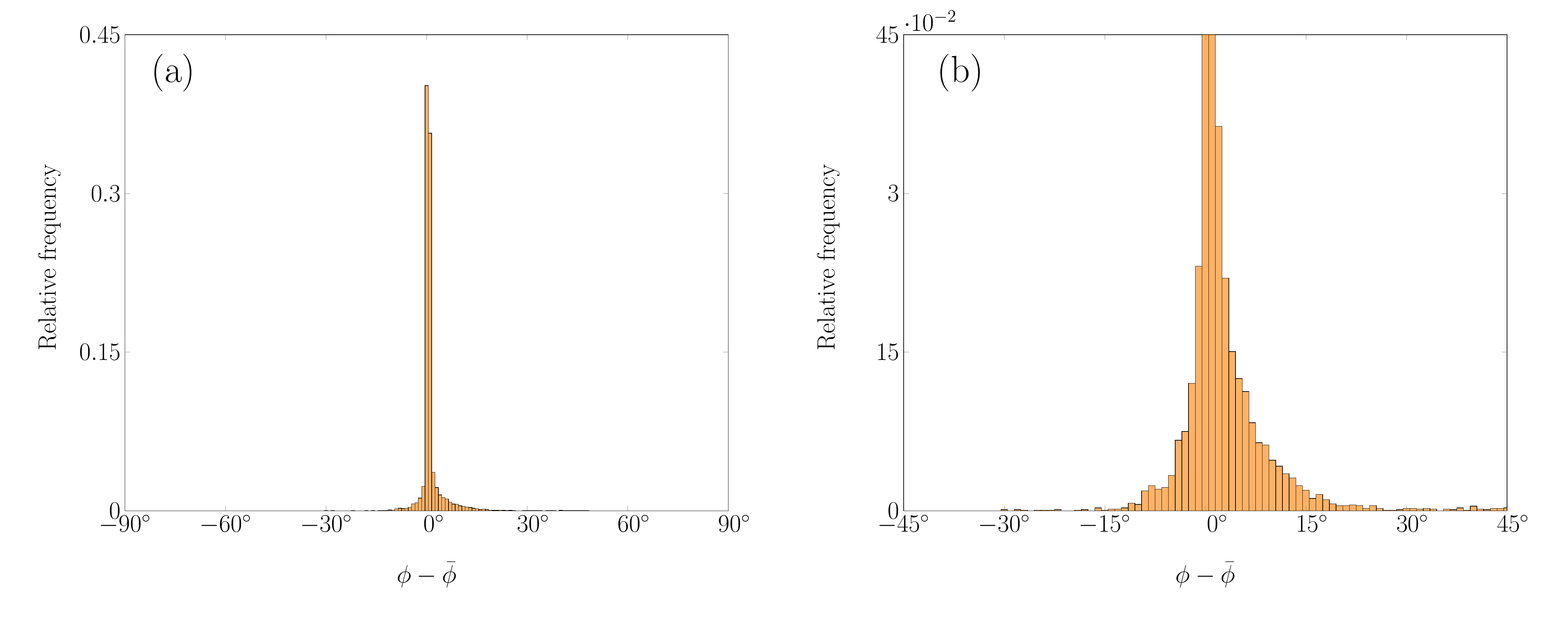}
	\caption{Relative frequency histograms of the error between the simulated M13 angles $ \phi $ and fit angles $ \bar{\phi} $, with $1^{\circ}$ bin widths, for the full analysis. Figure (a) shows the full error distribution with figure (b) zoomed in for clarity.}
	\label{fig:wssHist}	
\end{figure}

\begin{figure}[t]
	\centering
	\includegraphics[width=\textwidth]{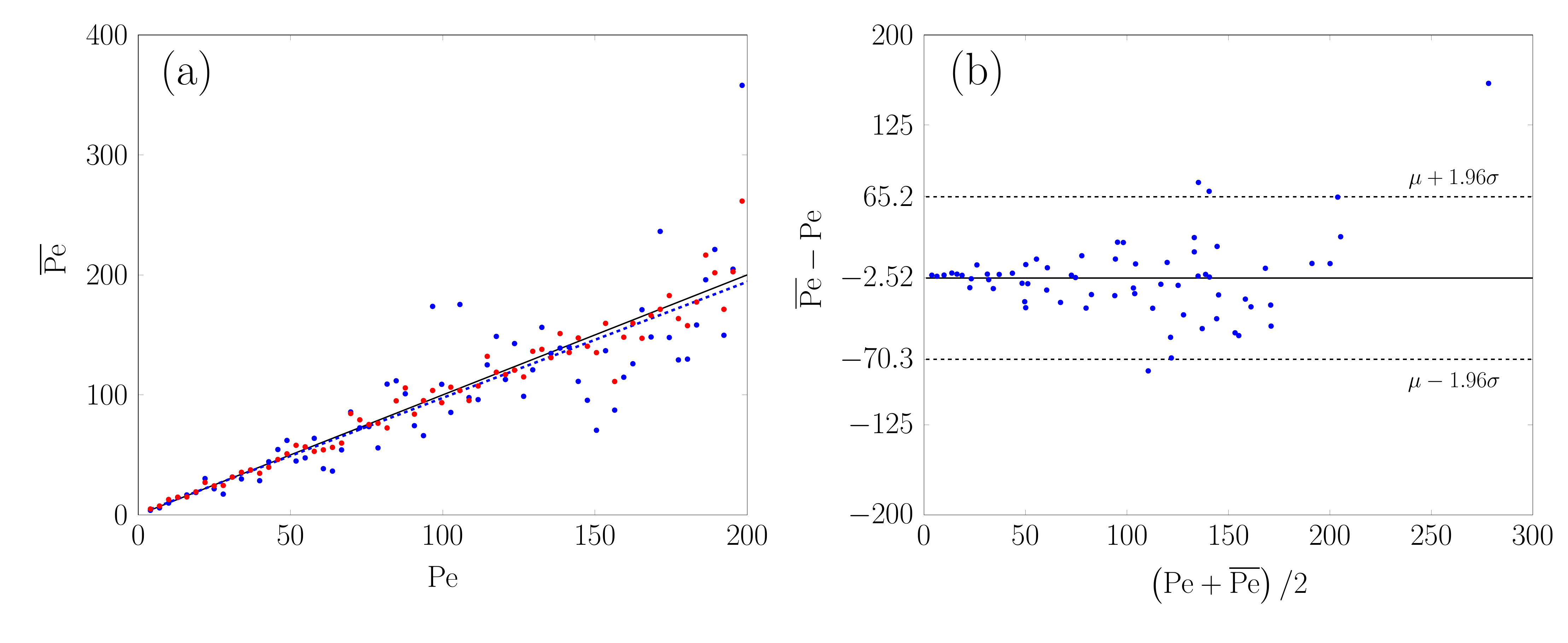}
	\caption{(a) plots the fit P\'eclet number $ \overline{\Pe} $ against simulated P\'eclet number $ \Pe $ for the full analysis. The blue dots show the fit after the image analysis has been carried out, while the red dots show the calculated P\'eclet number assuming the image analysis step is perfect. The black line shows what would be perfect correspondence between $ \Pe $ and $ \overline{\Pe} $, while the dotted blue line is the line of best fit to the data. (b) shows the Bland-Altman plot testing the fit data $ \overline{\mathrm{Pe}} $ against $ \Pe $. The solid lines show the mean of the difference between $ \overline{\Pe} $ and $ \Pe $, with the dotted lines being the $ 95\% $ confidence interval for the difference.}
	\label{fig:wssPe}	
\end{figure}

	Having shown that the error in the image processing step is well contained with greater than $ 90\% $ of fit angles deviating from the simulated angles by less than $ 5^\circ $, we move on to look at the error associated with the full analysis of WSS from a sample of orientation data. We do this in the same way as in section \ref{sec:case1}, however instead of using rejection sampling to obtain a sample with the required P\'eclet number, we take a single sample of $ 180 $ angles $ \phi $ from the marginal probability distribution $ \Phi $ at each $ \Pe $. This should give insight into the accuracy of the full analysis on experimental images, with additional error being introduced through the generation of the orientation sample. 
	
	The number of images successfully analysis, along with the number of fit orientation angles $ \bar{\phi} $ within $ 1^\circ $ and $ 5^\circ $ of simulated angles $ \phi $ are again shown in table \ref{tab:ipErr}, with the corresponding relative frequency histograms of the error between $ \phi $ and $ \bar{\phi} $ shown in figure \ref{fig:wssHist}. We see very similar results to that of Step 1, which is to be expected as we have not changed the image analysis portion of the methods, which is independent of angle distribution. 
	
	In figure \ref{fig:wssPe} we plot the P\'eclet numbers obtained from fitting the angles $ \bar{\phi} $. Included in this figure are the P\'eclet numbers given by fitting to the sampled angles (assuming a perfect image analysis method), where we can see the deviation about what would be perfect correspondence to the flow, which is a result of the restricted sample size in the simulations, and analogous to the error from having a restricted sample size in the related experiments. Here, the Bland-Altman plot, figure \ref{fig:wssPe}b, shows a mean difference between  $ \Pe $ and $ \overline{\Pe} $ of $ 2.52 $, with the standard deviation of the difference being $ 35 $. We see here that, despite this additional error, and the error from the image processing procedure in Step 1, we can reliably calculate the P\'eclet number relating to a given flow.

	\begin{figure}[t]
	\centering
	\includegraphics[width=0.9\textwidth]{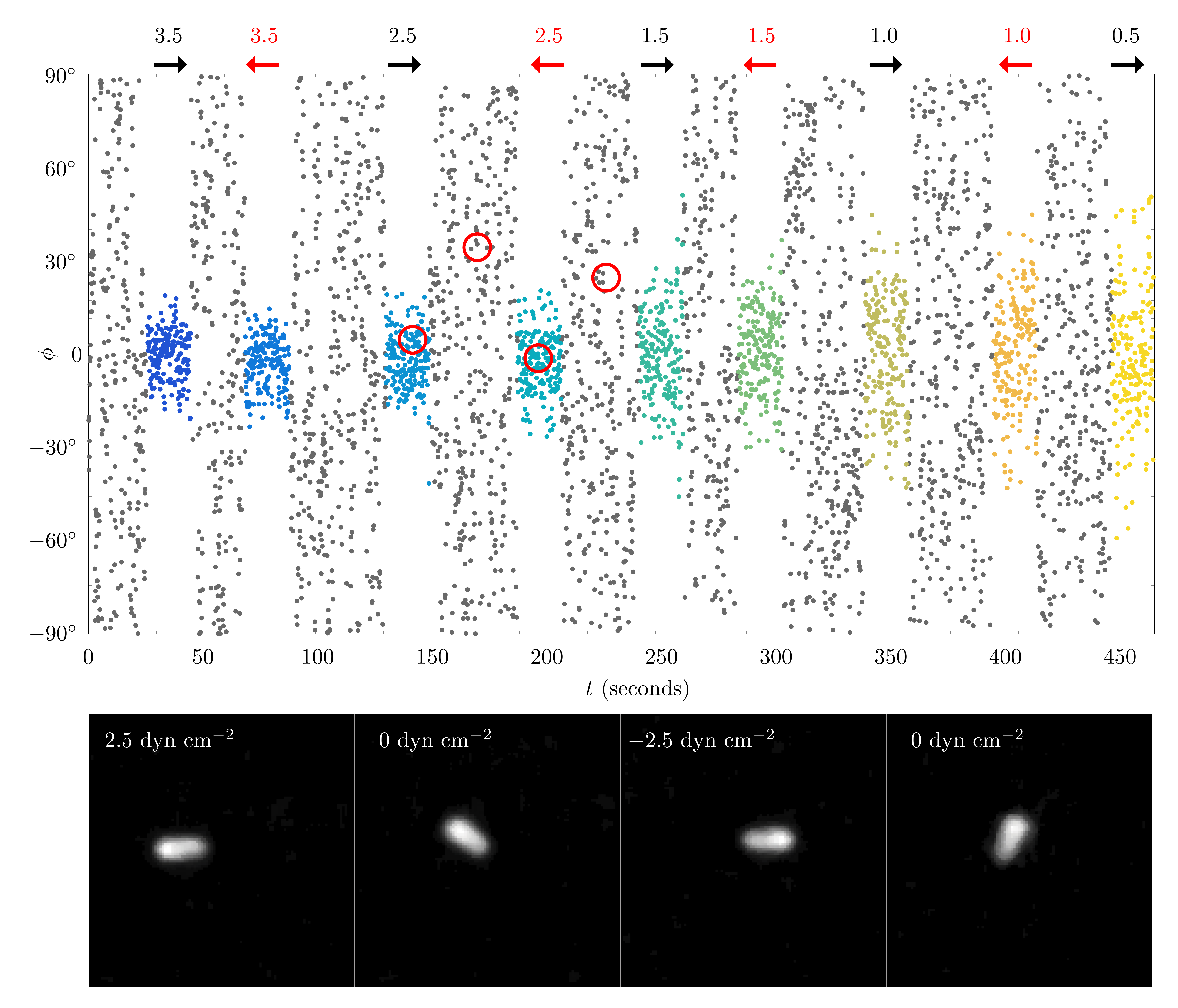}
	\caption{Plot showing the angles $ \phi $ obtained by fitting to the raw image data of Lobo et al \cite{lobo2015}. Here the new analysis data has been presented in the style of Lobo et al \cite{lobo2015} for ease of comparison. Each grey dot represents an image where there is no nominal WSS, while the nominal WSS for each other colour is written above the figure, with an arrow indicating the direction of flow. Shown below is a selection of experimental images for the circled frames.}
	\label{fig:phi}
\end{figure}
	
\begin{figure}[tp]
	\centering
	\includegraphics[width=\textwidth]{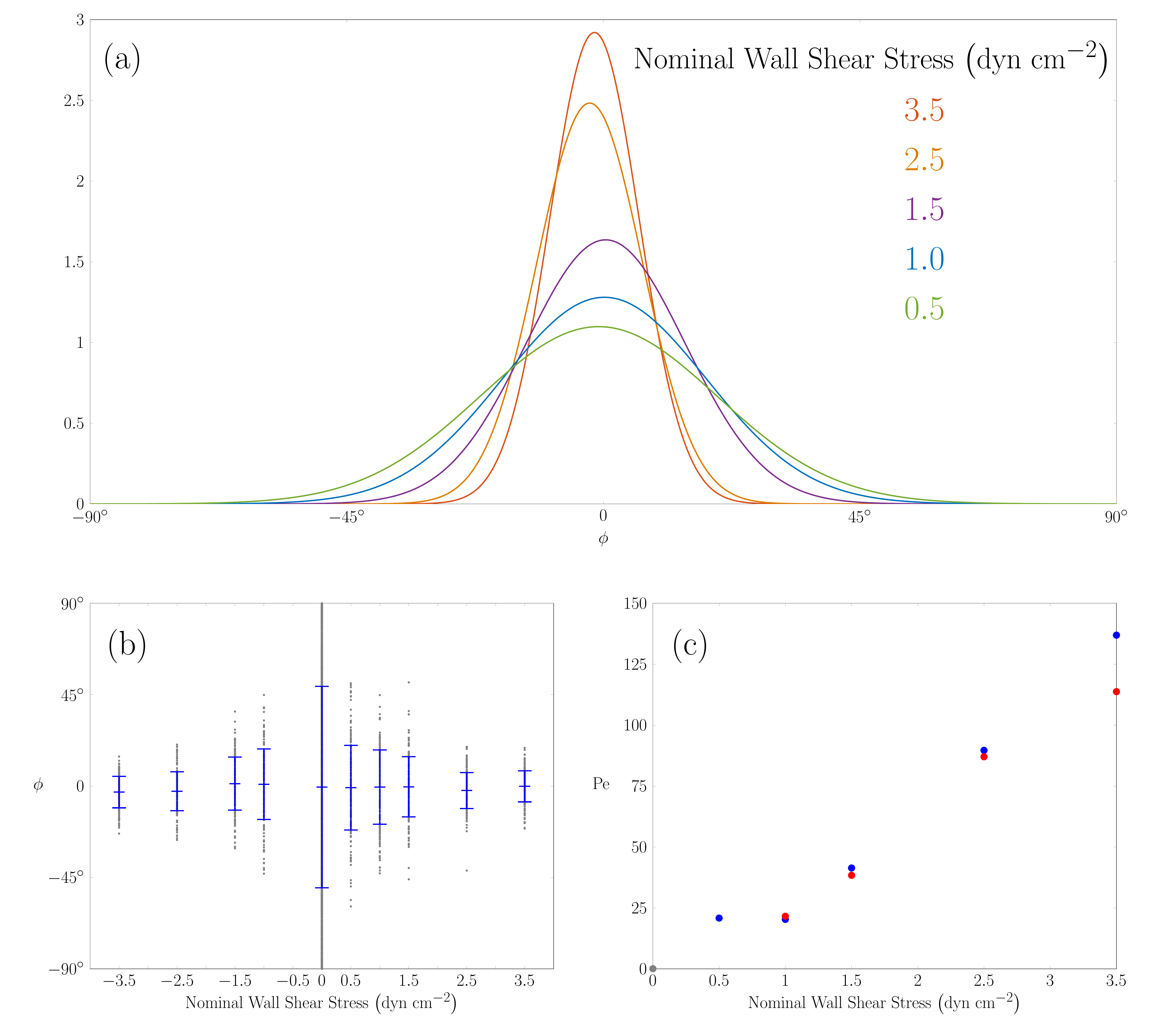}
	\caption{(a) Distribution of the angles $ \phi $, with lines showing a normalised Gaussian fit to the data. (b) Angle $ \phi $ against nominal WSS. Here the blue marks show the mean angle for each value of WSS, and the blue line show one standard deviation above and below the mean. (c) Here, the approximated P\'eclet number is shown against nominal WSS. The blue and red dots represent flow in the positive and negative directions respectively, while the grey dot represents the case of zero applied flow.}
	\label{fig:results}
\end{figure}


\section{Calculation of WSS from experimental image data}
\label{sec:results}

The angles $ \phi $ obtained from fitting the full series of raw image data from Lobo et al \cite{lobo2015} are shown in figure \ref{fig:phi}. Each point represents a single frame, with the corresponding applied nominal WSS, and direction, shown above the plot. Additionally, red circles show the location of four characteristic images, which are displayed at the bottom of the figure. It is clear by eye, before doing any in-depth analysis that, when flow is applied, there is a strong biasing of the distribution of the M13 angle $ \phi $ towards the direction of flow, and that this biasing effect is more pronounced the greater the nominal WSS. This is in agreement with the more detailed analysis shown in figures \ref{fig:results}a and \ref{fig:results}b. In figure \ref{fig:results}a we have fit a normalised Gaussian model to the data for individual flow rates, combining data from flows of the same magnitude in different directions. It is clear from these figures that as the nominal wall shear stress increases, the probability that the M13 is aligned with the flow (towards $ \phi = 0^{\circ} $) increases, with the standard deviation of the angles about $ \phi = 0^{\circ} $ decreasing. As expected the flow direction does not have an impact on the distribution of the M13, as can be seen in figure \ref{fig:results}b. Finally, we plot the estimated P\'eclet number for the flow in figure \ref{fig:results}c, where it is clear that with increased nominal wall shear stress, we have fit a larger P\'eclet number. We note that the P\'eclet number calculated for the $ 0.5 $ dyn cm$^{-2}$ flow appears to be larger than expected. We believe this to be due to the fact that the flow lies outside the sensitivity range of the M13 in the experiments; a longer M13 would have more sensitivity to lower levels of WSS. This assertion is discussed in more detail in section \ref{sec:conc}. We also see here that there is a slight discrepancy between the fit for the flows in the positive direction (blue) and negative direction (red). This is to be expected from the statistical nature of the fit owing to the Fokker-Planck model, and we also expect some difference due to the fact that the collagen IV surface is not completely flat leading to slight changes in flow behaviour in different directions. We believe that the fits in each direction are close enough to give credence to the viability of the fitting procedure.


\section{Conclusions}
\label{sec:conc}

It has recently been shown that a biological microrod (M13) can act as a wall shear stress sensor \cite{lobo2015} through flow-induced changes to its tethered Brownian motion. We have now developed and presented the first mechanistically rational analysis of this novel assay. This modelling and measurement framework consists of two steps, combining areas of mathematical modelling, fluid dynamics, and image analysis, namely

\begin{enumerate}
	\item \emph{Dynamics of a tethered Brownian fibre}
		
		Here, we have modelled the rotational Brownian dynamics of a tethered Brownian fibre system under homogeneous unidirectional shear flow. Given experimentally calculated orientation data for a M13 under flow, the modelled orientation probability distribution for the M13 allows the 
			calculation of a P\'eclet number for the flow, and hence a measure of the wall shear stress over a biologically relevant surface.
			
	\item \emph{Model-based image analysis}

		To complement the mathematical modelling of the Brownian dynamics, we have developed an rigorous and extensible framework for the analysis of a set of experimental images. We have tackled the inverse problem of image formation, the solution to which allows the accurate and reliable calculation of the M13 location in a heavily diffracted image. This framework allows the swift, accurate, and automated calculation of orientation data from experimental image data.

\end{enumerate}

We have applied this model to the problem of calculating wall shear stress, validating against the work of Lobo et al \cite{lobo2015}. This work differs from the previous analysis in that we have developed a principled and extensible framework for the analysis of the experimental data, as opposed to simply calibrating the method with the experimental results. In analysing the same data we have introduced the concepts of model-based image analysis and have tackled the inverse problem of image formation in order to locate the M13 in a series of experimental images. We believe that this approach to image analysis allows us to have more faith in the results, over more traditional image analysis techniques, due to the physics of image formation which have been included in the underpinning model, but also due to the statistical framework for modelling the Brownian motion of the M13 which enables multiple sources of error to be considered in the analysis. The techniques introduced here also offer the advantage of being completely automated once set up, there is no manual component unlike many other methods, which allows the analysis of much larger quantities of data than would have been previously possible. 

We have shown that the combination of the fluid dynamic modelling of a tethered M13, together with the model-based image analysis of the experimental images, can produce an estimated P\'eclet number for the flow, the ratio between shear-driven and Brownian-driven effects in the flow. Through simulations we have produced an estimation of the accuracy of the model, and have shown that this method can reliably produce biologically relevant results. The methods can also be tailored to detect particular types of flows. Rotational diffusion scales with length like $ \mathcal{D} \sim L^{-3} $, so small changes in M13 length have a large impact on rotational diffusion coefficient, and hence P\'eclet number. The impact of this is that M13 engineered to be slightly longer will have a smaller diffusion coefficient and hence enable the detection window to be extended to lower shear rates; slightly shorter M13 will have a larger diffusion coefficient, hence enabling the detection window to be extended to higher shear rates -- with the caveat that for orientation to be detected, diffraction associated with the emission wavelength places a lower limit on M13 length.

The theory in this paper provides methods for calculating the shear stress on a flat surface through imaging of a tethered M13. The extensibility of the presented framework means that only small modifications in the fluid dynamic modelling (section \ref{sec:pec}) are required in order to estimate the shear stress over more biologically relevant surfaces in vivo e.g.\ over the endothelial cell lining of a blood vessel. We would then be able to directly apply the methods for solving the inverse problem of image formation as discussed in section \ref{sec:meth}. Additionally, regarding the model-based image analysis, if we were able to accurately measure the optical diffusion in a given experimental set up, and relate this to the point spread function model \eqref{eqn:psf}, we should be able to obtain the full 3D reconstruction of the M13 location, which would then allow the use of the full probability density function $ \psi $, rather than the marginal PDF, $ \Phi $, as obtained in section \ref{sec:pec}. We would expect good results in the full 3D case, even if the surface is not perpendicular to the imaging plane provided there was some knowledge about the surface topography which could be taken into account in the fluid dynamics modelling. In addition such results could be improved through the use of multiple imaging planes to better capture the M13 position in full 3D space.

While in this work we have only considered the calculation of surface shear stress, the techniques developed here could have wider applications in the fields of micro-scale biology and image analysis. Of great interest is the application of the model-based image analysis techniques to experimental data of motile cells such as sperm. We believe that these techniques will be able to provide great insight into, for example, the measurement of sperm kinetics and morphology, and will have the potential for wide-ranging impact in fields such as fertility and animal husbandry.  

\section*{Data accessibility}

All data for the generation of figures can be found at \newline http://epapers.bham.ac.uk. All the code for this project can be accessed at https://github.com/meuriggallagher/phage.

\section*{Authors' contributions}

The contributions of the authors is as follows: M.T.G., D.J.S., and K.P.A. designed the research; all authors contributed to image pre-processing; M.T.G., D.J.S.,  and C.V.N. developed and implemented the mathematical models and methods, and analysed the data; M.T.G., and D.J.S. wrote the manuscript with additional input from K.P.A. and C.V.N.. All authors contributed intellectually to the work presented.

\section*{Acknowledgements}

We are grateful to the authors of Lobo et al \cite{lobo2015} for being forthcoming with the raw data and original methods. The expertise of Tim Dafforn, University of Birmingham, Alison Rodger, University of Warwick, and Matt Hicks, Linear Diagnostics Ltd., contributed significantly to the underlying research and the experimental data used in this report. The funding from a variety of sources to support this work is gratefully acknowledged: Engineering and Physical Sciences Research Council (Healthcare Technologies EP/N021096/1: D.J.S., M.T.G.); British Heart Foundation (Project Grant no. PG/15/37/31438: K.P.A.). K.P.A. also thanks Bizkaia talent (AYD-000-256) and the Medical Research Council (MR/P003214/1) for salary support.


\appendix
\section{Derivation of the advection-diffusion equation for a tethered fibre}
\label{app:der}

The flux of the probability density function $\psi\bra{\theta,\phi,t}$ is given by $\bm{J}=\psi \dirdot$, where $\dirdot(\theta,\phi)$ is the rate of change of $\bm{d}$ due to the combination of hydrodynamic and Brownian rotations. Denoting by $\angvel(\theta,\phi)$ the torque-free angular velocity of the particle induced by the shear flow, then the rate of change of $\dir$ under rigid body rotation is,
\begin{equation}
\dirdot = \angvel \times \dir.
\end{equation}
In the presence of the shear flow, a fibre rotating with angular velocity $\bm{\omega}$ is therefore given by,
\begin{equation}
\bm{T}^H=\mathcal{R} (\angvel - \bm{\omega})\mbox{,}
\end{equation}
where $\mathcal{R}$ is the rotational resistance matrix about the origin, taking into account the effect of the plane boundary. The $\theta$-dependence is a consequence of the boundary effect.

Following \cite{kim1991}, the Brownian torque on a suspension is given by,
\begin{equation}
\bm{T}^B=-\dir\times D(kT \log \psi) = -\frac{kT}{\psi} \dir\times \gradD  \psi\mbox{,}
\end{equation}
where $k$ is Boltzmann's constant and $T$ is absolute temperature. Torque balance $\bm{T}^H+\bm{T}^B=0$ then yields,
\begin{equation}
\mathcal{R}(\angvel - \bm{\omega}) - \frac{kT}{\psi} \dir \times \gradD \psi=0.
\end{equation}
Rearranging we have,
\begin{align}
\psi\dirdot &= \psi \angvel \times \dir \ - \mathcal{R}^{-1} \gradD \psi\mbox{,} \nonumber \\
            &= \psi \bm{\alpha} - \mathcal{D}\gradD \psi\mbox{,}
\end{align}
where $\mathcal{D}=\mathcal{R}^{-1}$ is the rotational diffusion matrix and $\bm{\alpha}=\angvel\times\dir$ is rotational advection vector.

\section{Numerical solution of the advection-diffusion equation}
\label{sec:appAdvDiff}
The diffusion tensor $\mathcal{D}$ for an axisymmetric body can be written,
\begin{equation}
\mathcal{D}=D_{\theta\theta}(\theta)\ethe\ethe+D_{\phi\phi}(\theta)\ephi\ephi
=
\begin{pmatrix}
D_{\theta\theta}(\theta) & 0            \\
0                        & D_{\phi\phi}(\theta)
\end{pmatrix}
\end{equation}
The advective term $\bm{\alpha}$ is given by,
\begin{equation}
\bm{\alpha}(\phi,\theta)=\alpha_\theta\ethe+\alpha_\phi\ephi=-\omega_\theta\ethe+\omega_\phi\ephi.
\end{equation}
where $\bm{\alpha}=\bm{\omega}\times\bm{d}$.

In component form, equation~\eqref{eqn:dimAdvDiff} can be written,
\begin{align}
&\mathrm{Pe} \left(\partial_\theta(\alpha_\theta\psi)+\cot\theta\alpha_\theta\psi+\frac{1}{\sin \theta}\partial_\phi\left(\alpha_\phi\psi\right)\right)\notag\\ 
&\hspace{1.25in}=\partial_\theta(\mathcal{D}_{\theta\theta}\partial_\theta\psi)+\cot\theta\mathcal{D}_{\theta\theta}\partial_\theta\psi+\frac{\mathcal{D}_{\phi\phi}}{\sin^2\theta}\partial_{\phi\phi}\psi.
\end{align}
where we have assumed that $ \psi $ is time independent, and dropped dashes on dimensionless variables for brevity. The system is solved numerically via a finite difference method to give an approximate solution $[\psi_{ij}]\approx \psi(\theta_i,\phi_j)$ ($i=1,\ldots,100$, and $j=1,\ldots,100$) on the domain $0^{\circ}< \theta<90^{\circ}$, $0^{\circ}\leq \phi<360^{\circ}$ for a given P\'eclet number $ \mathrm{Pe} $. 


\end{document}